\documentclass[a4paper]{article}
\usepackage{graphicx}

\usepackage[T2A,T1]{fontenc}
\usepackage[utf8]{inputenc}

\usepackage{apacite}
\usepackage{url} 

\begin{document}

%
%


\title{Identifying ground scatter and ionospheric scatter signals by using their fine structure at Ekaterinburg decameter coherent radar}

%
%




\author{I. A. Lavygin,V. P. Lebedev, K. V. Grkovich, O. I. Berngardt}

\maketitle


\begin{abstract}
The analysis of the scattered signal was carried out in the cases
of ground scatter and ionospheric scatter. The analysis is based on
the data of the decameter coherent EKB ISTP SB RAS radar. In the paper
the signals scattered in each sounding run were analyzed before their
statistical averaging. Based on the analysis, a model is constructed
for ionospheric scatter and ground scatter signal, based on previously studied mechanisms. Within
the framework of the Bayesian approach and based on large number
of the data, the technique for identifying the two types of signals
is constructed based on their different nature. 
The technique works without using traditional
SuperDARN methods for estimating scattered signal parameters - spectral width, 
Doppler drift velocity or ray-tracing.
The statistical analysis of the results was carried out.
The total error produced by our IQ algorithm over the selected data was $13.3\%$, that 
is about two times less than total error produced by traditional algorithms.

\end{abstract}

%
%
%


%
%

%


%
%
%
%

\section{Introduction}

One of the basic tools for studying ionospheric convection and magnetosphere - 
ionosphere interaction in high-latitude regions are 
SuperDARN (Super Dual Auroral Radar Network) radars \cite{Greenwald_1995, Chisham_2007} 
and similar pulse radars \cite{Berngardt_EPS}. 
The radars allow one to study radio signals scattered by small-scale irregularities and 
to use their dynamics for the ionospheric 
processes investigation \cite{Ruohoniemi_2005}.

The signal received by the radars consists of three basic parts -
noise of a different nature \cite{Berngardt_2018}, 
a signal scattered from ionospheric irregularities\cite{Ruohoniemi_1993} (ionospheric scatter, IS)
and a signal refracted in the ionosphere and scattered from the Earth's
surface \cite{Milan_1997} (ground scatter, GS). 
The physical mechanisms responsible for scattering from the Earth's surface and from ionospheric
irregularities are different, so the ways of interpreting the scattered signal are also different. 
However, the problem of identifying the signals scattered from ionosphere and
the signals scattered from the Earth's surface is of practical interest \cite{Blanchard_2009, Ribeiro_2011}. 

Ionospheric scatter has statistical nature and requires statistical averaging to estimate its parameters.
So one of the main techniques used for identifying ground scatter and ionospheric
scatter signals at the present time at these radars is the analysis
of average spectral characteristics of received signals \cite{Ponomarenko_2006}.
Usualy it is assumed that only ground scatter signal has sufficiently low Doppler 
shift and low spectral width \cite{Baker_1988,Ponomarenko_2006,Blanchard_2009}, other signals are ionospheric scatter. 
But in some cases this assumption is not correct. 
In the case when ionospheric irregularities are moving across the line-of-sight, the Doppler shift of ionospheric scatter 
is zero. This can lead to incorrect identification of the ionospheric scatter as ground scatter.
In the case when the background ionosphere is sufficiently disturbed, the Doppler shift
of ground scatter signal can be large enough  \cite{Hayashi_2010,Grocott_2013}. This can lead to incorrect 
identification of the ground scatter as ionospheric scatter.

Therefore various, more sophisticated methods are being developed
to solve the problem of identification of IS and GS signals: 
significant increase of spectral resolution by using longer sounding
sequences \cite{Berngardt_2015c}; complex spectral processing techniques 
\cite{Barthes_1998}; raytracing of radiosignal propagating
in the ionosphere \cite{Liu_2012}; complex spatio-temporal
analysis of the areas in which the scattered signal is observed \cite{Ribeiro_2011}.
However, the physical scattering models from the ground surface and ionospheric 
irregularities apparently were not taken into account.

In this paper we present a novel approach for identifying GS and IS signals 
without analysis of their average spectral characteristics. The analysis was made 
based on the EKB ISTP SB RAS radar data.
The method is based on the analysis of amplitude-phase structure of scattered signal in every single sounding. 
As a result a signal model was constructed for describing the scattered signal.
The use of this model allows us to identify GS and IS signals without calculating their average Doppler shift, 
average spectral width, or making additional qualitative considerations.

\section{EKB radar observations}

Ekaterinburg coherent decameter radar (EKB ISTP SB RAS) is a monostatic radar of CUTLASS
type developed by University of Leicester\cite{Lester_2004}.
It was assembled jointly with IGP UrB RAS under financial support
of the Siberian Branch of the Russian Academy of Sciences and Roshydrometeorological
Service of the Russian Federation at Arti observatory of IGP UrB RAS.
The radar antenna system is a linear phased array.
It provides a beam width of the $3^{o}-6^{o}$ depending on
sounding frequency and 16 fixed beam positions within $52^{o}$ field
of view. The spatial and temporal resolution of the radar is 15-45
km and 2 minutes, respectively. The radar frequency range 8-20 MHz 
allows the radar to operate in over-the-horizon mode.
The radar peak pulse power 10 kW allows it to operate up to 3500-4500
km radar range. Short sounding signals provide a low (about 600 Watt)
average transmit  power, this allows the radar to operate in a 24/7 monitoring
mode \cite{Berngardt_EPS}.
For regular sounding it uses complex multipulse sequences that provide high spatial and spectral 
resolution at the same time \cite{Farley_1972,Berngardt_2015c}.

\begin{figure}
\includegraphics[scale=0.3]{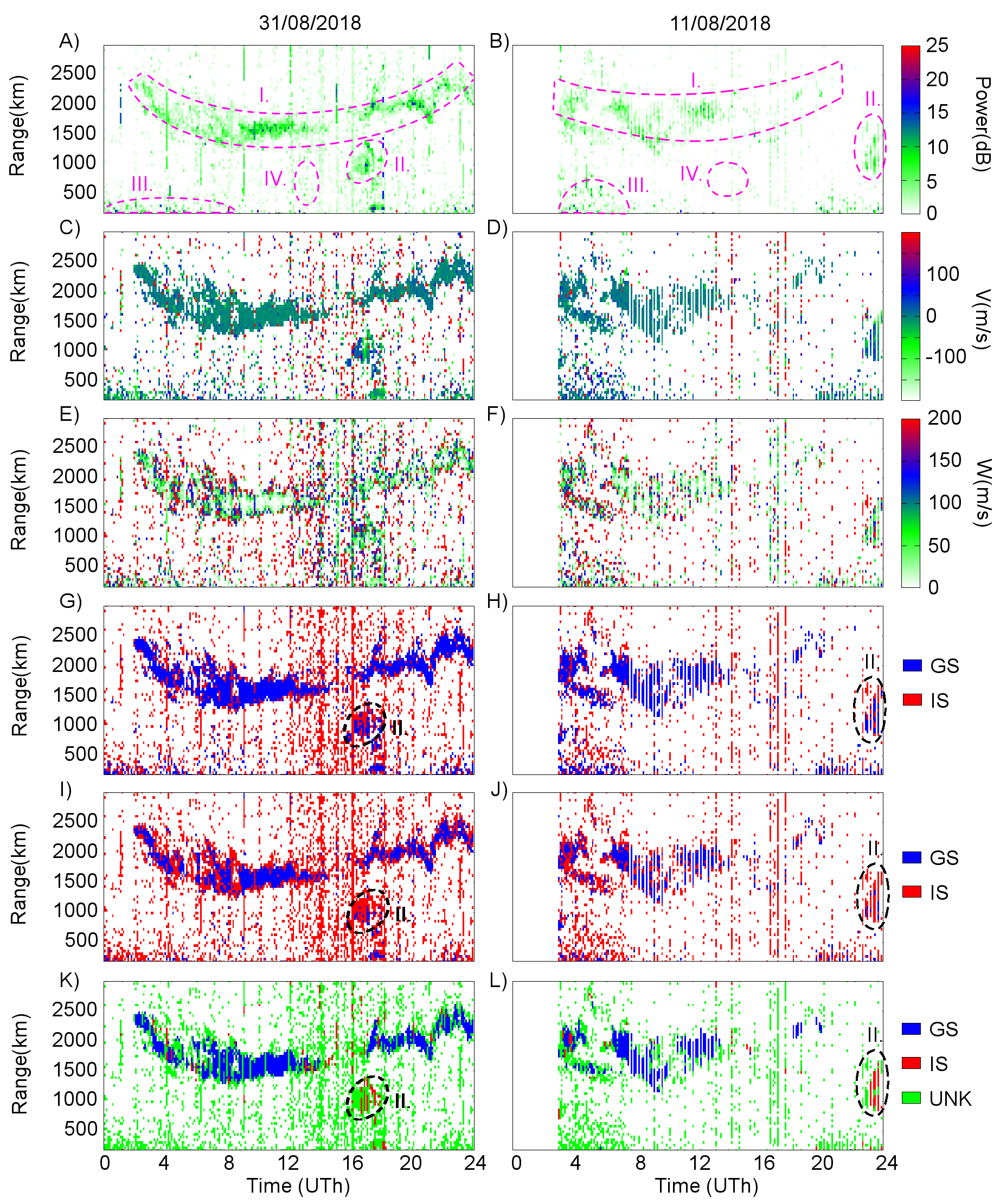}
\caption[test]{
An example of scattered signals received at EKB ISTP SB RAS radar
(at single beam). From the top to the bottom: A-B) is the power of the
scattered signal; C-D) is the Doppler velocity in the range -200 - 200
m/s; E-F) is the spectral width in equivalent velocity units; 
G-H) - signal types detected by \cite{Blanchard_2009} algorithm.
I-J) signal types detected by FitACF v.2.5 algorithm 
K-L) signal types detected by \cite{Ribeiro_2011} algorithm
(blue - ground scatter, red - ionospheric scatter, green - unknown type).
Regions marked by dashed lines are examples of:
I. - ground scatter; 
II. - ionospheric scatter; 
III. - near range echo and meteoric scatter; 
IV. - noise
}
\label{fig:FIG1}
\end{figure}

The basic technique of parameter estimation is the standard FITACF algorithm, 
developed and improved by the SuperDARN community \cite{Ponomarenko_2006, Ribeiro_2013}. 
The algorithm estimates the power, Doppler shift and spectral width of the scattered signal.
These parameters are estimated by fitting measured average signal autocorrelation function (ACF)
by two models: exponential and Gaussian ones\cite{Hanuise_1993}. The parameters can be interpreted in terms of 
scattering cross-section, line-of-sight drift velocity and lifetime of the ionospheric irregularities.
In Fig.\ref{fig:FIG1} are shown examples of the parameters obtained at EKB ISTP
SB RAS radar (at one of its beams). 

In Fig.\ref{fig:FIG1}A-B are shown the areas corresponding to the basic kinds of scattered signals: 
the signal scattered from the Earth's surface (GS, region I.),
the signal scattered from ionospheric irregularities (IS, region II.), 
scattering from meteor trails (meteor echo, region III.) and noise (region IV.).
Fig.\ref{fig:FIG1}A-F shows the basic characteristics
of the received signals: power (A-B), velocity (C-D) and spectral width(E-F).

The standard approach to identifying signals of different types is
based on the spectral parameters of the mean autocorrelation
function - the spectral width and Doppler frequency shift \cite{Ponomarenko_2006, Ribeiro_2013}. 
In this approach the GS signals correspond to small values of both
parameters, not exceeding some limit values \cite{Baker_1988,Ponomarenko_2006,Blanchard_2009}.

The latest version of FitACF \cite{FitACF_25} uses the following condition:

\begin{equation}
\frac{|v|}{\Delta v}+\frac{|w|}{\Delta w}<1
\label{eq:vw_gs_flag}
\end{equation}

where $\Delta v = 30m/s; \Delta w = 90m/s$; $v=\frac{c}{2}\frac{\delta f}{f_{0}}$ is Doppler drift velocity;
$w=\frac{c}{2}\frac{\sigma f}{f_{0}}$ is spectral width in velocity units; $f_{0}$ is sounding frequency.
Other combinations of the values correspond to IS signals. 

\cite{Blanchard_2009} suggest another GS condition:

\begin{equation}
|v|-(33.1[m/s]+0.139|w|-0.00133[s/m]w^{2})<0
\label{eq:vw_gs_flag2}
\end{equation}

\cite{Ribeiro_2011} suggest another algorithm for identifying GS. It is based on cluster analysis and depends 
on the cluster size (in the range-time coordinates) and average values of $v,w$ over the studied cluster. 
The algorithm is available at \cite{Ribeiro_Cluster}.

Fig.\ref{fig:FIG1}G-L shows the result of data identification by the above three algorithms. 
Red color corresponds to identifying the signal as IS, blue color corresponds to GS signals. 
Green color in Fig.\ref{fig:FIG1}G-L marks the areas that \cite{Ribeiro_2011} algorithm 
cannot confidently identify.
In Fig.\ref{fig:FIG1}G-L, region II one can see that there are cases when these different algorithms 
from the same initial values $v,w$ make different conclusions about the type of received signal.

As one can see from (\ref{eq:vw_gs_flag},\ref{eq:vw_gs_flag2}) the problem of identifying GS and IS 
is extremely important in the case when the ionospheric irregularities have a narrow spectrum and
a small Doppler shift, which is sometimes observed when the drift of the irregularities is 
perpendicular to the radar line-of-sight.
In these cases, the signal parameters take values close to the boundary for a particular algorithm, 
and therefore algorithms with different boundary values determine such signals as having different types. 
An example of such cases is shown in Fig.\ref{fig:FIG1}G-L, region II. As one can see, within a single 
area, which of qualitative considerations should consists of signals of the same type, 
the algorithms identify signals as having different types. 
Such an identification is apparently a fault, and one of the ways to solve this problem is to develop 
new approaches to identifying the type of scattering, that are not based on the Doppler shift or the 
spectral width of the received signal.

In the paper we investigate such a novel approach.
Currently most SuperDARN radars, as well as EKB ISTP SB RAS radar,
measure and store the full waveform of the scattered signal without its averaging over soundings.
The use of a full waveform of the scattered signal is useful for the
studies of meteor echo \cite{Yukimatu_2002}, for digital formation
of antenna pattern \cite{Parris_2008} and in some other applications.

Currently, only several SuperDARN radars have the capability
of sampling the scattered signal with high sampling rate, for example
\cite{Parris_2008}. 
Initially, the EKB radar did not have this capability, and digitized the 
received signal with low sampling frequency (one point per sounding pulse duration ($100-300\mu s$)).
To investigate fine structure of the scattered signals, the radar was reprogrammed
and sampling frequency was increased.
The maximal sampling frequency achieved by us in regular mode is 5 points per
sounding pulse duration. 
This mode corresponds to the range sampling rate $L_{d}=9km$ for the regular
spatial resolution $L_{p}=45km$.
EKB radar started its regular observations in this mode in February 2017.

To build the identification algorithm, we studied GS and IS signal properties separately.
From the obtained experimental data, we manually selected two test data sets.
In each data set there was only one intense response over the range - either
first-hop ground scatter or ionospheric scatter. In each selected set 
we have no significant doubt about the type of scattered signal.
For GS signal data set we selected the regions with a specific horseshoe-shaped
spatio-temporal dependence of power vs. range and time (See Fig.\ref{fig:FIG1}A-B, region I).
For the IS signal data set, we used mostly evening and night responses and some observations of daytime
ionospheric scatter clearly separated from the ground scatter (See Fig.\ref{fig:FIG1}A-B, region II).

\begin{figure}
\includegraphics[scale=0.5]{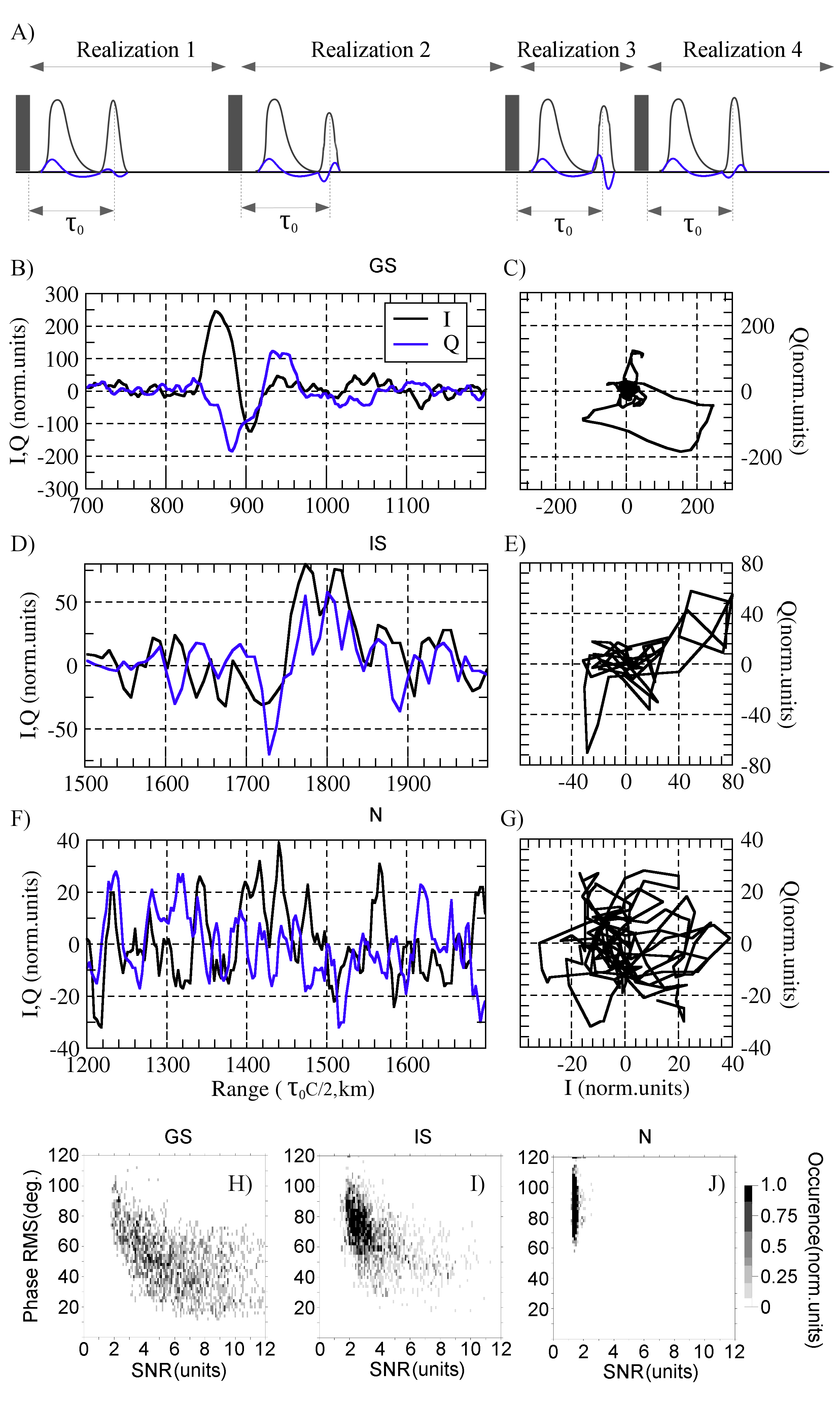}
\caption{
A) - sounding process, grey rectangles correspond to sounding pulses, black and blue lines 
correspond to I and Q components of the scattered signals;
B-G) example of realizations of the received signal (quadrature components
and phase of the signal) with a high sampling frequency in the cases
of the ground scatter (B-C), the ionospheric scatter (D-E) and the
noise (F-G). H-J) - distributions of detected elementary responses as a function
of SNR and the PRMS (\ref{eq:sqo_phase})
for different types of scattered signals: for ground scatter(H),
for ionospheric scatter(I), and for noise(J).}
\label{fig:FIG2}
\end{figure}

Fig.\ref{fig:FIG2}A illustrates the sounding process at the EKB radar. 
The sounding by non-equidistant sequence of sounding pulses produces 
a sequence of realizations of the scattered signal, the group delay of 
which $\tau_{0}$ is measured from the nearest sounding pulse. 
Below in the paper we analyze these realizations of the scattered signal, 
as a function of the realization number and group delay $\tau_{0}$.

Fig.\ref{fig:FIG2}B-G shows examples of the received signal fine structure
(amplitude and phase structure) in the cases of the ground scatter
(Fig.\ref{fig:FIG2}B-C), the ionospheric scatter (Fig.\ref{fig:FIG2}D-E)
and the noise (Fig.\ref{fig:FIG2}F-G), as a function of range $\frac{\tau_{0}c}{2}$. 
The ionospheric scatter and ground scatter are chosen with a large signal-to-noise ratio, so that
their fine structure is estimated accurately enough.
It can be seen from the Fig.\ref{fig:FIG2}B-G, that both types of
scattered signals (GS and IS) have a certain phase structure.
This allows us to consider them as 
signals having some structure. 
The noise (N) has nearly no phase structure. 
To identify the fine structure of IS and GS signals,
we carried out a detailed analysis of the scattered signals on a large
amount of data.

\section{Scattered signal model}
\subsection{Qualitative considerations}

Let us describe some known features of both signal types.

    The GS signal has been studied for a long time in various experiments.
It is known that this signal is formed by a strong refraction of
radiowaves in the ionosphere. The refraction leads to the focusing of radiowave
at the boundary of the dead zone (skip distance) (a zone at which the receiving of
radiowaves at a given frequency is impossible) \cite{Budden_1985}.
Scattering of this high-amplitude signal by the ground surface irregularities
causes a strong received signal at a range corresponding to the
boundary of the dead zone. The irregularities of the Earth's 
surface with scales of the order of the wavelength (tens of meters)
are nearly static (excluding sea surface).
The background ionosphere at the scales of the Fresnel zone radius
(of the order several kilometers) is responsible for focusing the
radio signal and also varies relatively slow from realization to realization. Therefore, the GS
signal in the first approximation is a signal with a static phase-amplitude
shape. The signal arises at a constant delay (radar range). 
Its initial phase can vary from realization to realization. 

The shape of ground scatter is asymmetric relative to the position of its
maximal energy, and this correlates well with experimental observations 
\cite{Bliokh_1987}. 
Due to its static shape, the GS signal shape can be detected by coherent accumulation over realizations.

A traditional
approach to interpret IS signal is the statistical model of large
number of random scatterers \cite{Farley_1969, Rytov_1988, Ishimaru_1999}. 
The model corresponds well to the experimental
data \cite{Farley_1969, Moorcroft_1987, Andre_1999, Moorcroft_2004}
and can be used to producing realistic simulators of the received
data \cite{Ribeiro_2013b}, and to develop signal processing techniques
for signals accumulated over small number of soundings (realizations)
\cite{Reimer_2016}.

An ionospheric scatter model without averaging was suggested in 
\cite{Grkovich_2011}. It was shown that in VHF band 
the scattered signal in most cases can be interpreted as a superposition of small 
number of equivalent elementary responses.
Each response repeats the sounding signal shape and differs from realization
to realization by amplitude, initial phase, and Doppler shift. 
Below we follow the model \cite{Grkovich_2011} and suggest that 
elementary response shape for ionospheric scatter can be detected by coherent 
accumulation of the signal over realizations. Let us explain why it is possible.

We expect that $300\mu s$ (sounding pulse duration) phase changes of ionospheric scatter are small.
Actually, the Doppler drift velocities for field-aligned ionospheric irregularities are defined by electric 
fields and usually do not exceed 1-2~km/s \cite{Greenwald_1995}. For radar sounding frequency 10-11MHz the $300\mu s$ phase variations 
do not exceed $10-20^{o}$. So for the used
sounding frequencies and sounding pulse duration the difference in Doppler shift of each elementary response 
should not significantly change elementary response phase shape.

We can also assume that regions of the most powerful (effective) ionospheric scattering have nearly static positions 
so we can use coherent accumulation.
Actually, it is known that the most powerful ionospheric scattering from field-aligned irregularities comes from the 
regions of effective scattering where radiowave propagation trajectory is nearly orthogonal to 
the Earth magnetic field \cite{Greenwald_1995}. 
So, from qualitative point of view, the regions of effective scattering are controlled by large-scale structure of ionospheric electron density
and by structure of the Earth magnetic field. During several tens of milliseconds (corresponding to characteristic repetition 
period of sounding pulses and consequently to minimal lifetime of elementary response that can be detected) both variate slow.
Within the quantitative approach \cite{Berngardt_2016}, the regions of effective scattering 
are also the regions with a given level of refraction and with satisfied aspect sensitivity 
conditions. So, the positions of the regions should not change during several tens of milliseconds.
Thus, within the framework of the model \cite{Grkovich_2011} the IS 
signal in several consequent realizations can be approximated by a superposition of several effective responses, 
that does not change their position (range) and shape from realization to realization.
All the responses have similar phase structure, but different amplitude and initial phase.

As a result, the signal model, that describes in the first approximation
both GS and IS signals, is a superposition of independent elementary responses. 
During analysis time each elementary response is characterized by:
\begin{itemize}
\item an unknown amplitude-phase shape as a function of range that does not change from realization to realization;
\item an unknown position (range) that does not change from realization to realization;
\item an unknown initial phase varying from realization to realization;
\item an unknown amplitude varying from realization to realization;
\item an unknown model lifetime (lifetime of elementary response), during which the model can be used.
\end{itemize}
As it will be shown later, the amplitude-phase shapes (fine structure) and lifetimes of elementary responses in GS and IS signals are 
different and can be used for identification of the signal types.

\subsection{Fine structure detection algorithm}

To determine the shape of elementary response for
both kinds of signals (GS and IS), we use the coherent accumulation
technique.

At the first stage we estimate the position (radar range or radar delay $\tau_{o}$, 
see Fig.\ref{fig:FIG2}A) of the most powerful elementary response.
Within a gate of ranges (about 300~km, the value is related with minimal delay between sounding pulses $~2.4 msec$) 
we look for the range, providing maximal average signal-to-noise ratio (SNR). 
At this stage we average SNR over 30 sounding sequences or about 1.5 second.  The averaging of SNR is made only over signals,
scattered from the first pulse in each sequence. It is necessary to skip interference effects, important for other sounding 
pulses in the sounding sequence.
Later we will study only the elementary responses with average SNR $>2$.

At the second stage we estimate initial phases of elementary responses.
To estimate the unknown phases of $N$ elementary responses in $N$ consequent realizations, 
the model phase of each single elementary response $\phi_{i}^{M}\left(t,k,\psi_{i}\right)$ 
is supposed to be linear:

\begin{equation}
\phi_{i}^{M}\left(t,k,\psi_{i}\right)=k\cdot(t-\tau_{o})+\psi_{i}
\label{eq:lin_phase}
\end{equation}
where $i$ is realization number, and $k,\psi_{i}$ are unknown parameters.

As it was already mentioned, even the large Doppler shifts observed in the
ionosphere lead only to a slight phase changes at the duration of 
the sounding pulse. Therefore, the use of the linear model
(\ref{eq:lin_phase}) for the phase of elementary scattering response
should be sufficient. The parameter $k$ is the phase distortion
factor caused by Doppler shift of ionospheric scatter and 
is assumed to be the same for all consequent realizations.
The parameter $\psi_{i}$ is the initial phase of the scattered signal
and varies from realization to realization.
The vector of model parameters $(k,\psi_{0},\psi_{1},....\psi_{N})$ defined in (\ref{eq:lin_phase}) 
is determined based on the minimization for the root mean square
deviation of the phase (PRMS) $\Omega$:

\begin{equation}
\Omega=\sqrt{\frac{1}{N_{r}T_{p}}\sum_{i=1}^{N_{r}}\int_{\tau_{0}-\frac{T_{p}}{2}}^{\tau_{0}+\frac{T_{p}}{2}}\left(\phi_{i}(t)-\phi_{i}^{M}\left(t,k,\psi_{i}\right)\right)^{2}dt}=min
\label{eq:sqo_phase}
\end{equation}

The minimization of (\ref{eq:sqo_phase}) is made over the region 
limited by the duration of the sounding pulse $T_{p}$
with the center at the delay $\tau_{o}$, calculated at previous stage.
The model phase (\ref{eq:lin_phase}) is linear over all the parameters, so 
the problem (\ref{eq:sqo_phase}) reduces to a system of linear equations and can be solved analytically.
The value of minimal PRMS (\ref{eq:sqo_phase}) 
is used to verify the adequacy of the phase structure model (\ref{eq:lin_phase}).

For study we use data obtained in 8 days during autumn, spring and winter seasons.
Fig.\ref{fig:FIG2}H-J shows the distributions of detected elementary responses,
as a function of their SNR and their 
PRMS (\ref{eq:sqo_phase}). Fig.\ref{fig:FIG2}H corresponds to processing
GS signals, Fig.\ref{fig:FIG2}I  corresponds to processing IS signals, 
and  Fig.\ref{fig:FIG2}J  corresponds to processing noise signals. 
In Fig.\ref{fig:FIG2}J one can see that elementary responses of the noise are characterized
by low SNR about 1.5 and by PRMS over $90^{o}$.
This relates to quasi-random nature of the noise and its nearly constant
amplitude over the range. Fig.\ref{fig:FIG2}H-J shows that the distributions of elementary responses 
in IS and GS are very close, but SNR of IS is lower than SNR of GS.

To compute the elementary response shape $U(t)$ all the $N$ realizations 
are rotated by their initial phases $\psi_{i}$ and averaged. 
So the signal accumulation in the region of duration $T_p$ near the region 
of maximal SNR is made coherently:

\begin{equation}
U(t)=\frac{1}{N}\sum_{i=1}^{N}u_{i}(t)e^{-i\psi_{i}}
\label{eq:model_acc}
\end{equation}

\begin{figure}
\includegraphics[scale=0.5]{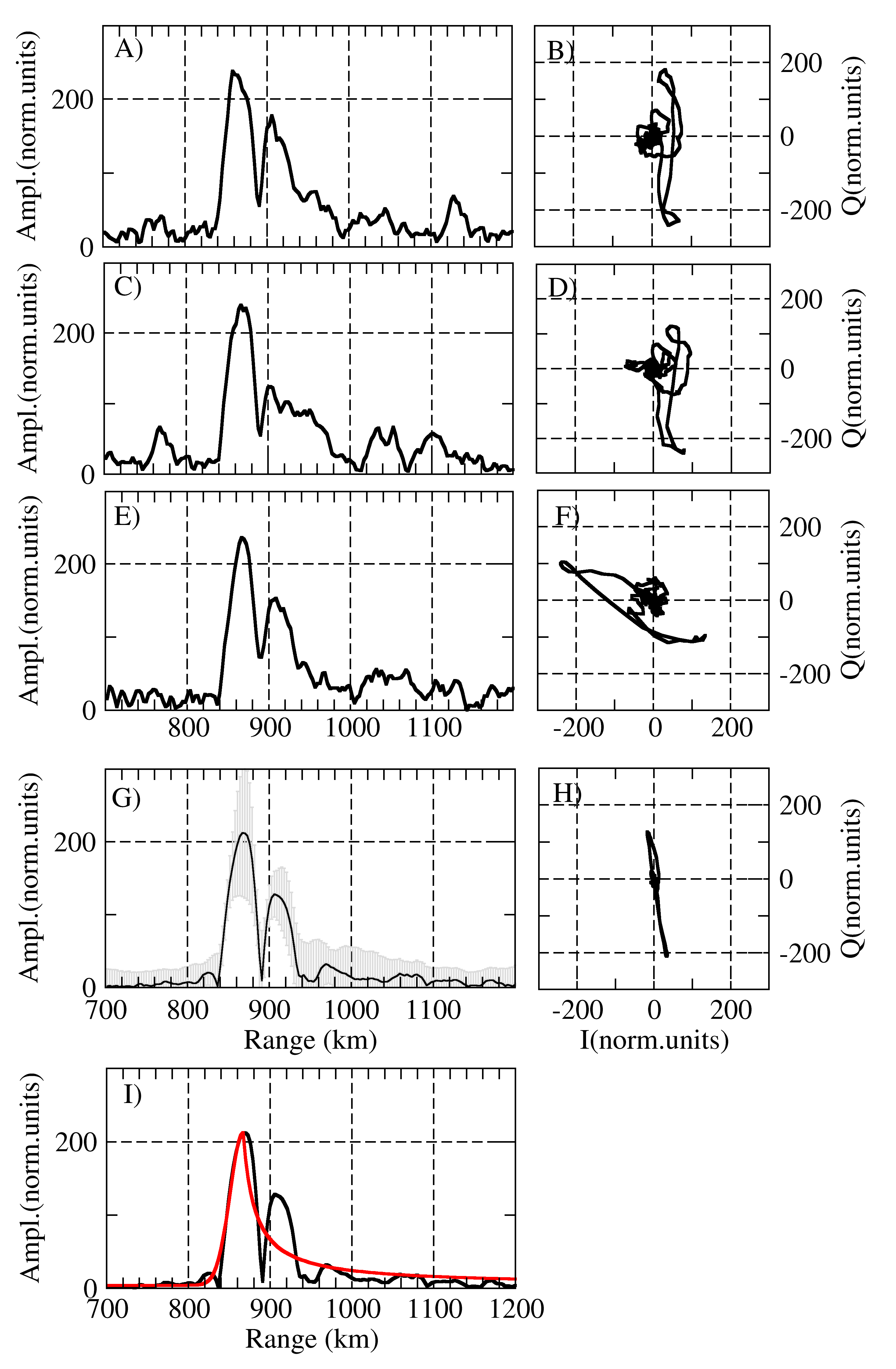}
\caption{Ground  scatter signals - amplitude (A,C,E,G,I) and phase diagrams
(B,D,F,H). A-F) 3 consequent realizations in the group G-H) Elementary 
response calculated over 30 sounding pulses (\textasciitilde{}300msec). Grey color at G) corresponds to RMS during accumulation.
The red line at I) shows the approximation of the accumulated signal by
the model (\ref{eq:a_model}).}
\label{fig:FIG4}
\end{figure}

\begin{figure}
\includegraphics[scale=0.5]{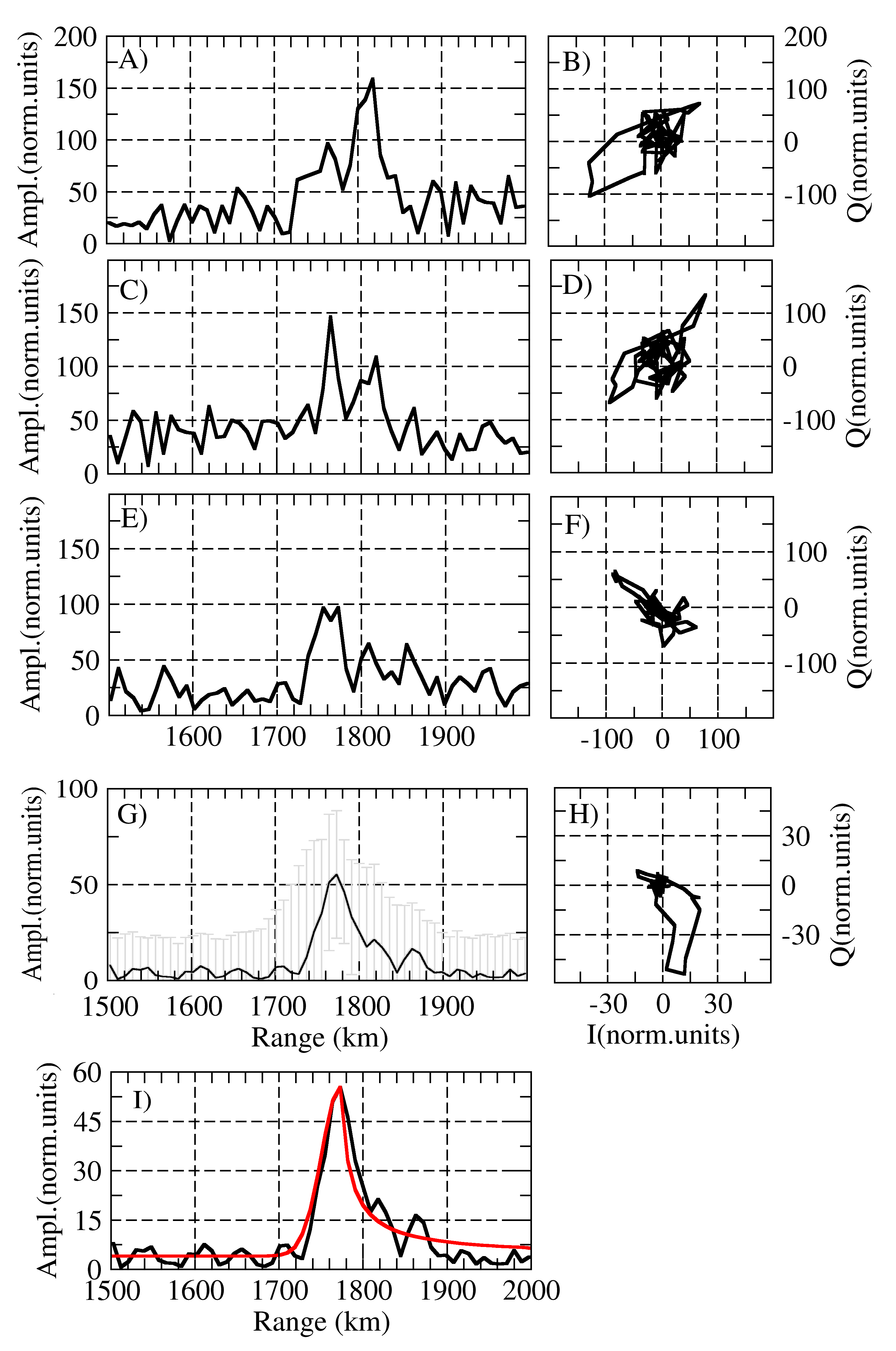}
\caption{Ionospheric scatter signals - amplitude (A,C,E,G,I) and phase diagrams
(B,D,F,H). A-F) 3 consequent realizations in the group G-H) Elementary response
calculated over 30 sounding pulses (\textasciitilde{}300 msec). Grey color at G) corresponds to RMS during accumulation.
The red line at I) shows the approximation of the accumulated signal by
the model (\ref{eq:a_model}).}
\label{fig:FIG5}
\end{figure}

\begin{figure}
\includegraphics[scale=0.5]{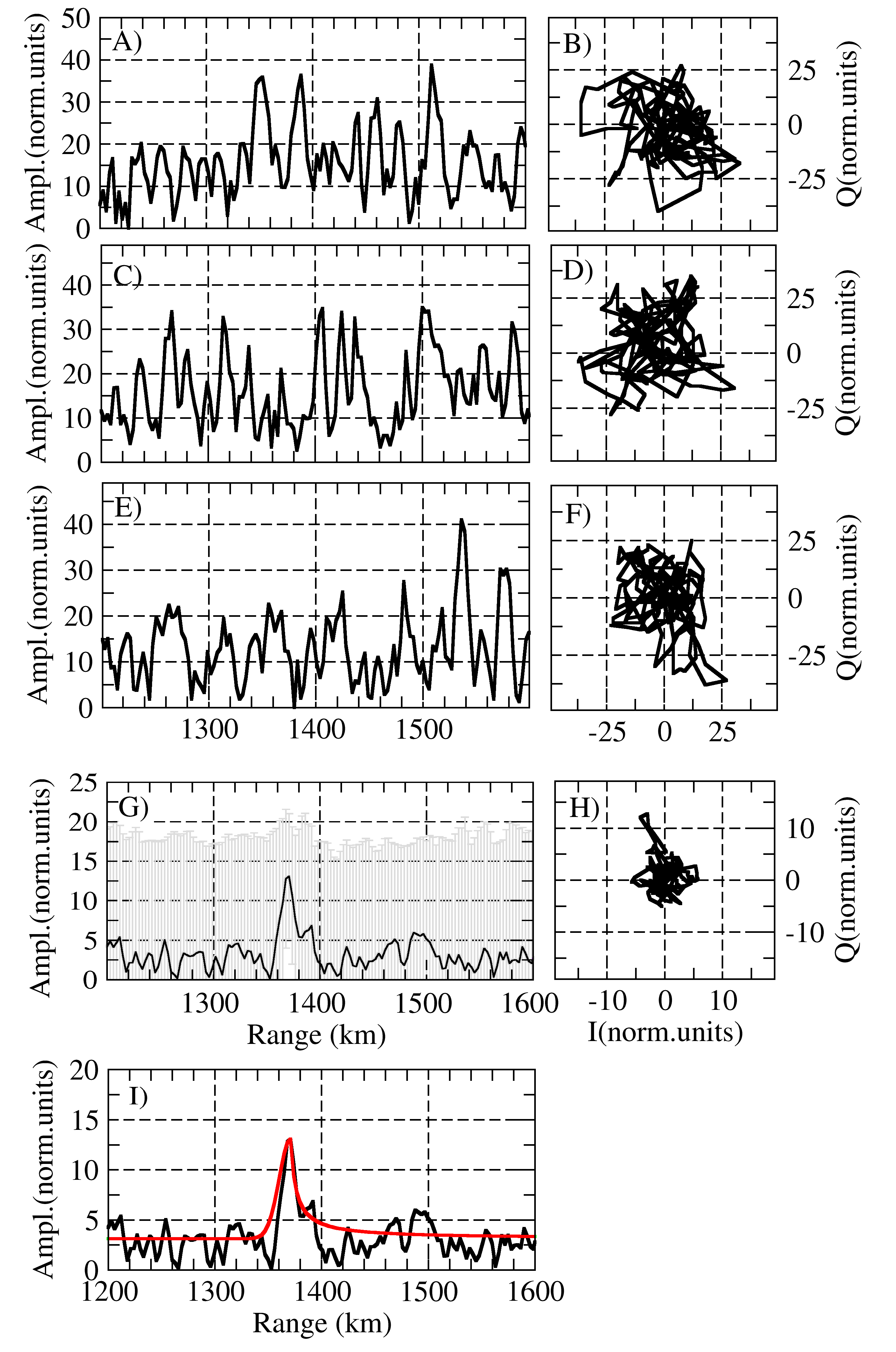}
\caption{Noise signals - amplitude (A,C,E,G,I) and phase diagrams (B,D,F,H).
A-F) 3 consequent realizations in the group G-H) Elementary response calculated
over 30 sounding pulses (\textasciitilde{}300 msec). Grey color at G) corresponds to RMS during accumulation.
The red
line line at I) shows the approximation of the accumulated signal by the model
(\ref{eq:a_model}).}
\label{fig:FIG6}
\end{figure}

In Fig.\ref{fig:FIG4}A-H, Fig.\ref{fig:FIG5}A-H and Fig.\ref{fig:FIG6}A-H
are shown examples of elementary responses in GS signals, 
IS signals and noise signals and RMS of accumulated signal
(Fig.\ref{fig:FIG4}G, Fig.\ref{fig:FIG5}G and Fig.\ref{fig:FIG6}G), as well as examples of signals 
in sequent realizations.

It can be seen from the figures that elementary responses in the GS
signals, in contrast to the IS signals
and noise, have essentially asymmetric shape and longer back
edge than the front one.

\subsection{Parameters of elementary response shape}

Let us check the regularity of the observed asymmetry over the experimental data.
To do this we use the following unified model that can approximate elementary responses both in the GS signals 
and in the IS signals:

\begin{equation}
a(t,\{A,B,C,D\})=D+
\left\{
\begin{array}{l}
Ae^{-\left(\frac{^{t-\tau_{0}}}{C}\right)^{2}};  t<\tau_{0}\\
\frac{A}{1+(t-\tau_{0})/B};  t>\tau_{0}
\end{array}
\right.
\label{eq:a_model}
\end{equation}

where $\tau_{0}$ is the radar delay to the maximal amplitude
of the scattered signal; $A$ is the maximal amplitude of the accumulated
signal; $B$ and $C$ are the parameters to be estimated, that characterize
the duration of the left and right edges, respectively; $D$ is the noise level.

The choice of this model can be qualitatively justified by the shape of the asymptotic
solution for the GS power, characterized by a sharp front edge, and
by the smooth back edge \cite{Tinin_1983, Bliokh_1987}. 
The comb amplitude structure of the accumulated signal, as well as its phase structure has 
not been investigated in the paper. 

Parameters $A,B,C,D$ are estimated by fitting the elementary response 
by the model (\ref{eq:a_model}). 
The model is linear over the $A,D$ and nonlinear over $B,C$.
So $A,D$ can be calculated analytically. To speed up the search for $B$ and $C$ 
we use an integral approach: the integrals of the elementary response on the left and right of $\tau_{0}$ 
should be equal to the corresponding integrals of the model function:

\begin{equation}
\left\{
\begin{array}{l}
\int_{\tau_{0}}^{\infty}(a(t,\{A,B,C,D\})-D)dt=\int_{\tau_{0}}^{\infty}(U(t)-D)dt \\
\\
\int_{0}^{\tau_{0}}(a(t,\{A,B,C,D\})-D)dt=\int_{0}^{\tau_{0}}(U(t)-D)dt
\end{array}
\right.
\end{equation}

The duration of left and right edges $T_{R},T_{L}$ of elementary response are calculated 
from the model parameters $B,C$.
To determine them 
we search for the points at which the model function (\ref{eq:a_model}) value
becomes equal to the given threshold level $\varepsilon_{L,R}$:

\begin{equation}
\left\{
\begin{array}{l}
a(t-T_{L},\{A,B,C,D\})=\varepsilon_{L}A+D \\
a(t+T_{R},\{A,B,C,D\})=\varepsilon_{R}A+D
\end{array}
\right.
\end{equation}

When the model (\ref{eq:a_model}) is fitted into a real sounding 
signal of $300\mu s$ duration, each of the calculated edge 
durations $T_{L},T_{R}$ should be equal to $150\mu s$ 
(half of the sounding pulse duration). 
Thus we determine the values of these threshold 
levels: $\varepsilon_{L}=0.5, \varepsilon_{R}=0.2$. 
The obtained values of $T_{L},T_{R}$ are used to
estimate duration of the elementary response edges directly in kilometers or microseconds.
Fitting examples are shown in Fig.\ref{fig:FIG4}I,\ref{fig:FIG5}I,\ref{fig:FIG6}I by red line.

By using the technique described above, we calculated the statistical
distributions of the duration of the left and right edges of elementary responses in IS
and GS signals over the experimental data set.
The results are shown in  Fig.\ref{fig:FIG8}A-B.
One can see that the characteristics of the elementary responses 
in different signal types are different: the elementary responses in
GS signals (Fig.\ref{fig:FIG8}A) are more asymmetric
and have smoother right edge in comparison with the right edge of the
elementary responses in IS signals (Fig.\ref{fig:FIG8}B). 
At the same time,
the elementary response in IS signal has relatively symmetrical
edges. This does not contradict the previously described models for
GS\cite{Tinin_1983} and IS \cite{Grkovich_2011} signals, and validates
the use of these models in the problem under consideration.

\subsection{Elementary response lifetimes}

Let us investigate the characteristic lifetimes of elementary responses. 
We estimate the lifetime by analysing the normalized cross-correlation coefficient between two different
realizations, as a function of the delay between them. Following
the approach described above the calculation of the correlation coefficient
is made over a region of maximal signal-to-noise ratio near
the delay $\tau_{0}$ (see Fig.\ref{fig:FIG2}A). The duration of the region is determined
by expected duration of the right and left edges $T_{R,0},T_{L,0}$.
The correlation coefficient becomes: 

\begin{equation}
R(i)=max_{n}\left\{ R_{n,n+i}=\frac{\int_{\tau_{0}-T_{L,0}}^{\tau_{0}+T_{R,0}}u_{n}(\tau)u_{n+i}^{*}(\tau)d\tau}{\sqrt{\int_{\tau_{0}-T_{L,0}}^{\tau_{0}+T_{R,0}}|u_{n}(\tau)|^{2}d\tau\int_{\tau_{0}-T_{L,0}}^{\tau_{0}+T_{R,0}}|u_{n+i}(\tau)|^{2}d\tau}}\right\} 
\label{eq:K_corr}
\end{equation}

where $u_{n}^{*}$ is the complex conjugate value of the signal $u_{n}$
received in n-th realization.

\begin{figure}
\includegraphics[scale=0.4]{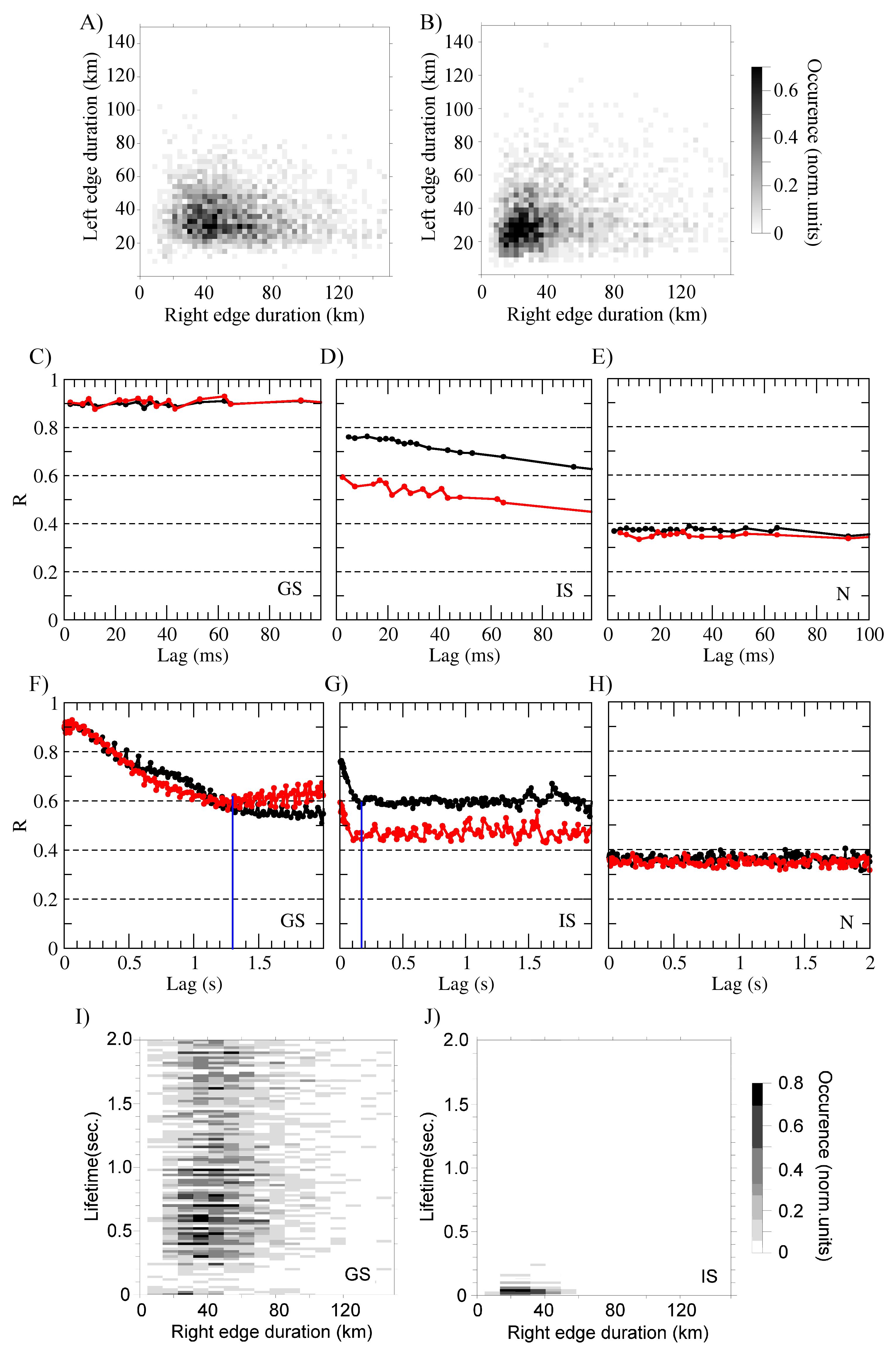}
\caption{
A-B) are the distributions of edges duration of the received
signals: A) is the distribution for GS; B) is the distribution
for IS.
C-H) - correlation coefficient for different signal types. C-E) are
is correlation coefficient at small lags comparable with sounding
sequence duration for elementary response in GS (C), in IS (D),
and in noise (E) signals; F-H) is correlation coefficient at large lags comparable
with averaging interval in regular sounding mode, for elementary response in GS
(F), in IS (G) and in noise (H) signals. Blue vertical
line at (D-E) corresponds to elementary response lifetime. Black and red lines at 
A-F correspond to different experiments; I-J) - distributions
of elementary responses over their lifetime (in seconds) and the right
edge duration (in km.) for GS (I) and for IS (J) signals.}
\label{fig:FIG8}
\end{figure}

As one can see in Fig.\ref{fig:FIG4},\ref{fig:FIG8}A, the duration
of ground  scatter signal in a single realization (before its coherent
accumulation) can reach up to 200-300 km. At the same time, the duration
of the left edge for both kinds of signals usually does not exceed
60 km ($400\mu s$, see Fig.\ref{fig:FIG8}A-B).
Therefore, we choose the following values for the duration of the right
and left edges used for calculation of correlation coefficient (\ref{eq:K_corr})
$T_{R,0}=400\mu s$, $T_{L,0}=1600\mu s$.

For a detailed analysis of the elementary response lifetime we developed
an algorithm, that works for arbitrary
lags, both smaller and exceeding the duration of the whole sounding
sequence (70 ms). 
The main property of
the sounding sequences is related with the properties of Golomb rulers \cite{Berngardt_2015c}:
the combinational lags between different sounding pulses are
always different and practically uniformly cover the region of lags
within the duration of the sounding sequence. 
So for delays less than sounding sequence duration we calculate
the correlation coefficient between elementary responses separated by
delays (lags) corresponding to the combinational lags between the sounding pulses. 
This approach looks close to calculating autocorrelation function (ACF) 
by standard SuperDARN processing technique.
 To evaluate the correlation coefficient at large lags exceeding the duration 
of the sounding sequence we calculate it at lags corresponding to the delay between
the first sounding pulse of each sounding sequence and the pulses of all
subsequent sounding sequences. 
Then we average correlation coefficient at each obtained lag over the available pairs with identical lag between them. 
Analysis of the correlation coefficient at 
so large lags traditionally is not carried out at SuperDARN
radars and at EKB radar. Most often this is associated with the complexity
of total synchronization of all sounding sequences. 
The described algorithm allows us to obtain
the correlation function of elementary responses at lags both less and greater than sounding sequence length.

Examples of the obtained correlation function for different kinds of scattered
signals are shown in Fig.\ref{fig:FIG8}C-H. It can be seen from the
Fig.\ref{fig:FIG8}C-E that the elementary responses in GS and IS signals
differ significantly from the elementary responses in the noise - they have higher correlation
coefficient at small lags. 
In Fig.\ref{fig:FIG8}D,G it is shown that the correlation coefficient of elementary responses in IS signals
increases at small lags, and this allows us to interpret the IS signal
as a result of scattering by scatterers with a relatively short lifetime
(hundreds of milliseconds). 
When the SNR decreases,  the  feature still persists, 
although it becomes less pronounced,  and
the maximum correlation coefficient at small lags decreases.
This does not contradict with field-aligned 
irregularities lifetime estimated in other experiments \cite{VILLAIN_1996}.

Fig.\ref{fig:FIG8}C,F show that
correlation coefficient of elementary responses in GS signals also decreases
with a lag, but the characteristic rate is much lower and elementary 
response lifetime in GS is longer than in IS. 
From Fig.\ref{fig:FIG8}C-D one can see, that in some cases it is difficult to differ GS
from IS by lifetime  using lags, provided by standard sounding sequences. 
This corresponds to small IS spectral widths case.

The analysis of correlation at extra large lags, compared with whole averaging
interval for regular sounding, is shown in Fig.\ref{fig:FIG8}F-H.
It can be seen from Fig.\ref{fig:FIG8}F-H that 
the average lifetime of elementary responses in GS signals ($>$1s)
exceeds the lifetime of elementary responses in IS signals($<$250ms).
This corresponds well with the physics of the signal formation. 
GS signal
is a signal scattered by nearly stationary surface inhomogeneties.
Its nonstationarity is mainly related with the existence of medium-
and large-scale ionospheric irregularities. They affect the refraction
of this signal and cause its decorrelation with time. 
The lifetime of individual small-scale irregularities (causing IS signals) is short.
It can be estimated from the spectral width of the IS signals ($>4Hz$,$>40m/s$), and therefore
rarely exceeds $250ms$. Thus, the measured elementary response lifetime does
not contradict with known characteristics of IS signals.

In Fig.\ref{fig:FIG8}F-G one can see that correlation coefficient falls 
with delay ('lag') and reaches a certain stable ('noise') level.
So we define the elementary response lifetime as the delay at which the correlation 
coefficient becomes lower than certain threshold level $R_{th}$. 
This level is calculated over the lags larger than 2 sec:

\begin{equation}
\begin{array}{l}
R_{th}=<R>+\Delta R \\
\\
<R>=\frac{1}{T_2-T_1} \int_{T_1 = 2sec}^{T_2 > T_1} R(t) dt \\
\\
\Delta R= \sqrt{ \frac{1}{T_{2}-T_{1}} \int_{T_1 = 2sec}^{T_2 > T_1} (R(t)-<R>)^2 dt}
\end{array}
\end{equation}

This approach allows us to automatically calculate the elementary response lifetime.

\section{Identification of IS and GS signals}

It was shown above that elementary response in GS and IS signals have different
shape and different lifetime. This allows us
to construct effective technique for identifying these signals by
their amplitude-phase and correlation characteristics.

One of the standard approaches to signal identification is the method
for testing statistical hypotheses \cite{Lehman_2005}. This method
reduces the problem of separating signals to the problem of determining
the detection boundary shape in the multidimensional space of signal
characteristics. The signals inside the boundary are identified
as IS signals, and the signals outside the boundary are identified
as GS signals. There are several methods of making such a boundary
shape, and they are based on minimizing the sum of the errors of the
first and second kind (errors of incorrect acceptance and incorrect
rejection of the hypothesis). We used the simplest Bayesian inference,
under assumption of the equal probability of IS and GS signals.

To calculate boundary for identification algorithm, we used the distribution of elementary response
characteristics in a three-dimensional parameter space (the duration
of the right edge, the duration of the left edge, and the lifetime). 
In Fig.\ref{fig:FIG8}I-J) it is shown that the
distribution of elementary response parameters for IS signal lies within a bounded region in the parameter
space. It is bounded by a certain surface around the coordinates center (small
lifetimes, short right edges). The parameters of elementary responses in GS signal
are outside the region (large lifetimes, long right
edges). In the first approximation, there is no significant correlation
between the duration of the right edge and the lifetime
(Fig.\ref{fig:FIG8}I-J), as well as between the left and right edges (Fig.\ref{fig:FIG8}A-B). 
So we use an ellipsoid with the axes along the coordinate
axes as separation boundary:

\begin{equation}
\frac{x^{2}}{a^{2}}+\frac{y^{2}}{b^{2}}+\frac{z^{2}}{c^{2}}=1
\label{eq:elips}
\end{equation}

The coordinates x, y and z are the elementary response lifetime,
the duration of the left edge and the duration of the right edge of
the elementary response, respectively. The ellipsoid axis
sizes - $a,b,c$ are to be determined.
Elementary responses with parameters inside this boundary (\ref{eq:elips}) correspond to IS signals, other correspond to GS signals.

To determine the parameters of the boundary, three-dimensional discrete distributions of elementary response parameters
$P_{IS}(x,y,z)$ and $P_{GS}(x,y,z)$ were constructed from two experimental
data sets (for GS and for IS). The step of the discretization of distributions
over the lifetime was 2.5~ms, and over
the left and right edges - 3~km. The search for
the optimal values of the parameters $a,b,c$ was made numerically,
by a direct search over the grid. The optimum condition
was the Bayesian criterion minimizing the sum of errors of the first
and second kind:

\begin{equation}
\Omega=\int_{r\left(\theta,\alpha\right)>\rho_{d}\left(\theta,\alpha\right)}P_{IS}(x,y,z)dxdydz+\int_{r\left(\theta,\alpha\right)<\rho_{d}\left(\theta,\alpha\right)}P_{GS}(x,y,z)dxdydz=min
\label{eq:elips_omega}
\end{equation}

Using polar coordinates $\left(r,\theta,\alpha\right)$ allows us to reduce
the separation problem to checking the conditions $r\left(\theta,\alpha\right)>\rho_{d}\left(\theta,\alpha\right)$
and $r\left(\theta,\alpha\right)<\rho_{d}\left(\theta,\alpha\right)$,
where $\rho_{d}\left(\theta,\alpha\right)$ is the equation of the
ellipsoid surface in polar coordinates.
Integration in the Cartesian coordinate system is made over the region
outside the surface of the ellipsoid in the first term, and over the
region inside this ellipsoid - in the second term. 
Search for optimum (\ref{eq:elips_omega}) over the data set with more than 13 thousand realizations
gives the following separation boundary (\ref{eq:elips}) parameters: 
a = 285 ms, b = 120 km, c = 429 km.

\begin{figure}
\includegraphics[scale=0.08]{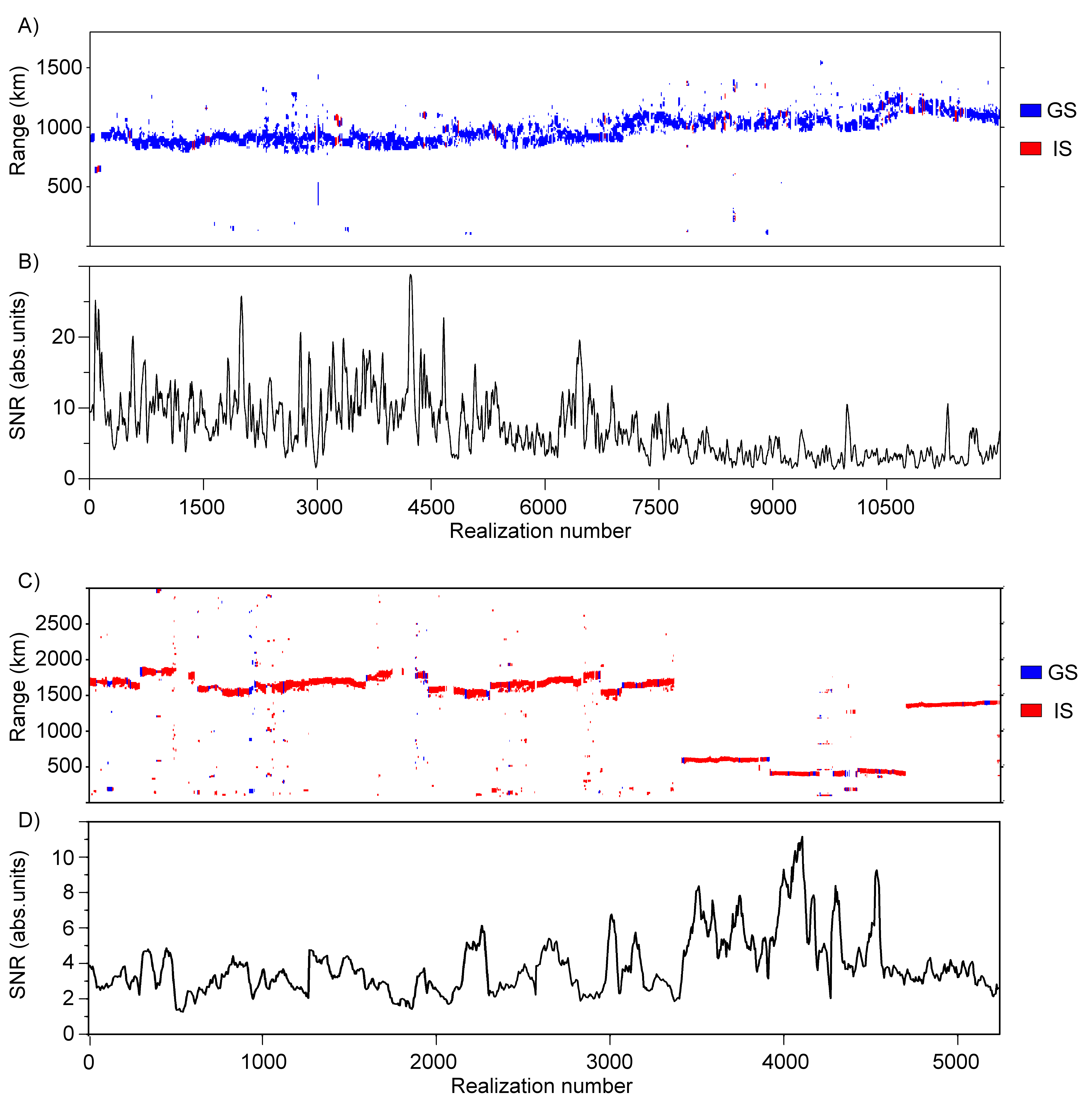}
\caption{
The results of identifying signal type by the new identification algorithm. 
At (A,C) blue dots mark the signals, identified as GS, red dots correspond to identification as IS signals.
A) - results of processing GS signals; B) - peak SNR of GS signals; C) - results of 
processing IS signals; D) is the peak SNR of IS signals.
}
\label{fig:FIG10}
\end{figure}

We use about 19 thousand realizations (13 thousand from training data set and 6 thousand from testing 
data set) to verify the technique. The results are shown in Fig.\ref{fig:FIG10}. As our analysis has shown, the 
accuracy of GS identification is 95.1\%. Accuracy of IS identification is 88.6\%. The total 
identification error is 16.3\%.

Fig.\ref{fig:FIG10} shows an examples of identification of  
the signal type for GS (Fig.\ref{fig:FIG10}A) and IS (Fig.\ref{fig:FIG10}C) signals.
As one can see in Fig.\ref{fig:FIG10}A,C, in most cases the algorithm works correctly 
and it does not depend on the SNR (Fig.\ref{fig:FIG10}B,D).
This is an indirect sign of the validity of the developed model and the identification
technique.


\section{Non-local character of the algorithm}

Most of the used signal GS/IS identification algorithms are local ones: the signal type is determined only by its local characteristics, 
measured at a given spatio-temporal point. 
In the case of simple algorithms, for example, \cite{FitACF_25,Blanchard_2009}, the radius of non-locality is constant and corresponds to 
spatial resolution (15-45 km) and temporal resolution (2-6 seconds) of measurements.
In more complex algorithms, like \cite{Ribeiro_2011}, it is not constant and depends on the characteristics of the scattered signal and 
the scattering region.

Our algorithm is a non-local one. The algorithm estimates the type of signal not at the selected range, but in the range area.
The size of this area is determined by the amplitude-phase
structure of elementary response.
To investigate the non-locality of the algorithm, we made a numerical simulation. We modeled simultaneous presence of two 
signals of different types, at different distances and with different amplitudes. This allows us to study the dependence of 
algorithm detection errors in different situations.
We carried out numerical modeling in three variants of the problem: to determine the parameters of signals that have equal 
amplitudes; to identify the signals in case IS is more powerful than GS; to identify the signals in case GS is more powerful than IS.

\begin{figure}
\includegraphics[scale=0.25]{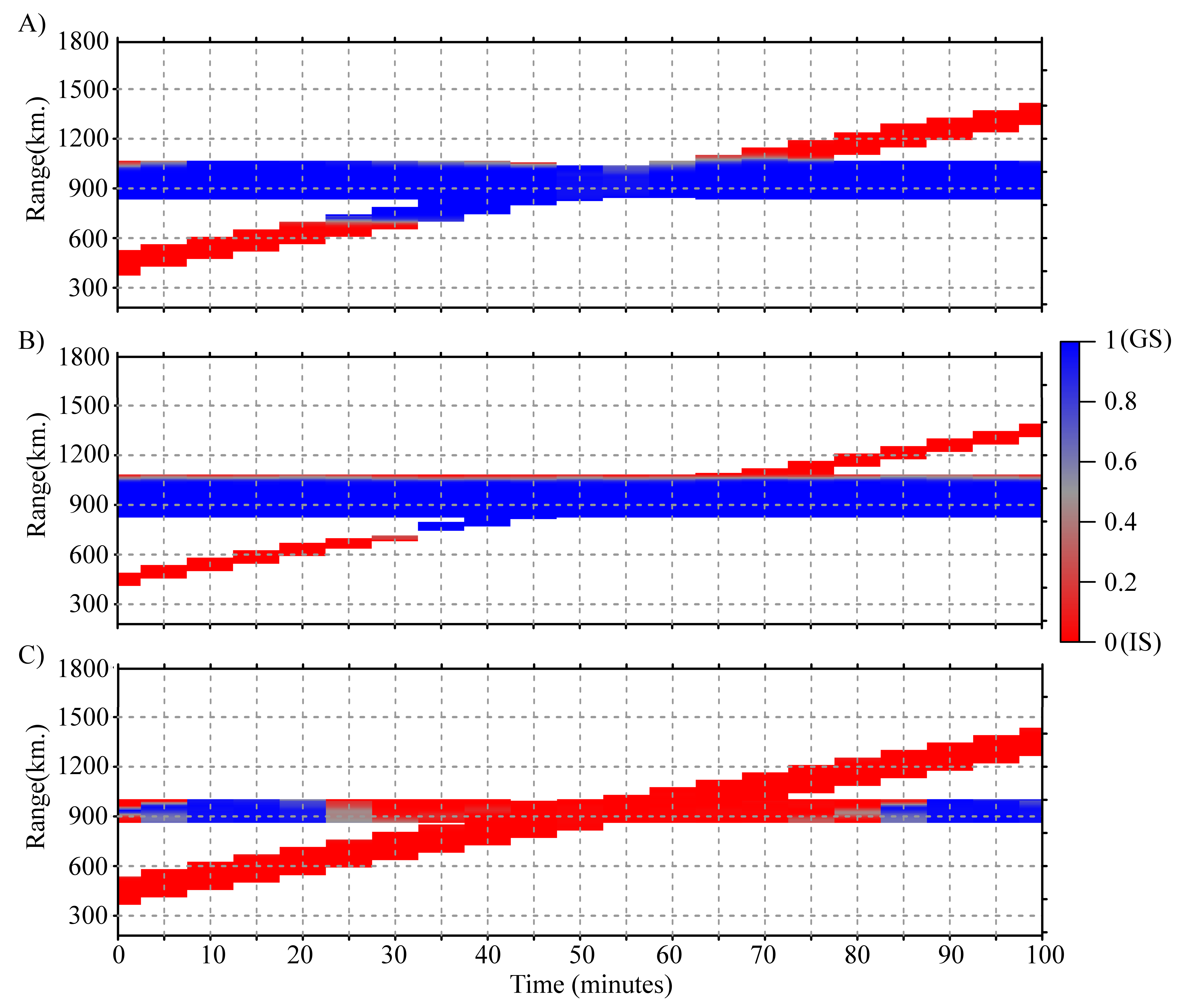}
\caption{
The results of the identification in a simulated case of two overlapping signals of different type. 
A) The signals have equal power B) GS signal is more powerful than IS; C) IS signal is more powerful than GS.
}
\label{fig:FIG13}
\end{figure}

The simulation results are shown in Fig.\ref{fig:FIG13}A-C. It can be seen that the algorithm begins to mix the characteristics 
of two signals when they are observed simultaneously in  the analysis window (when the distance between them $<200km$, Fig.\ref{fig:FIG13} A). 
In this case, the identified type of signal is inherited from the more powerful signal (Fig.\ref{fig:FIG13}B,C). In case of signals of 
equal power, IS signal identification dominates (Fig.\ref{fig:FIG13}A). 
The change in the width of the detected areas in Fig.\ref{fig:FIG13}A-C is related to the change in the SNR, which affects the amount of data processed.

\section{Comparison with other classification techniques}

We have compared the effectiveness of our IQ-method with the techniques developed earlier 
\cite{FitACF_25, Blanchard_2009, Ribeiro_2011} in their standard configuration described 
in the corresponding papers and available programs.
To compare these techniques, we manually selected from the available EKB radar data 80 observation 
periods of GS observations and 51  periods of IS observations. In each of the 
selected periods, the nature of the signal according to expert judgment was not in doubt. The data 
used for the comparison is shown in Supporting Materials, areas are marked with closed regions. 
Data with SNR$>2$ were processed by all the mentioned identification algorithms.

For the analysis, data obtained in the following 8 days of 2017 and 2018 was used:
21/10/2017, 09/12/2017, 25/12/2017, 09/03/2018, 15/03/2018, 26/04/2018, 27/04/2018, 05/05/2018.
The total number of analyzed cases of scattering was about 15 thousand.

\begin{figure}
\includegraphics[scale=0.3]{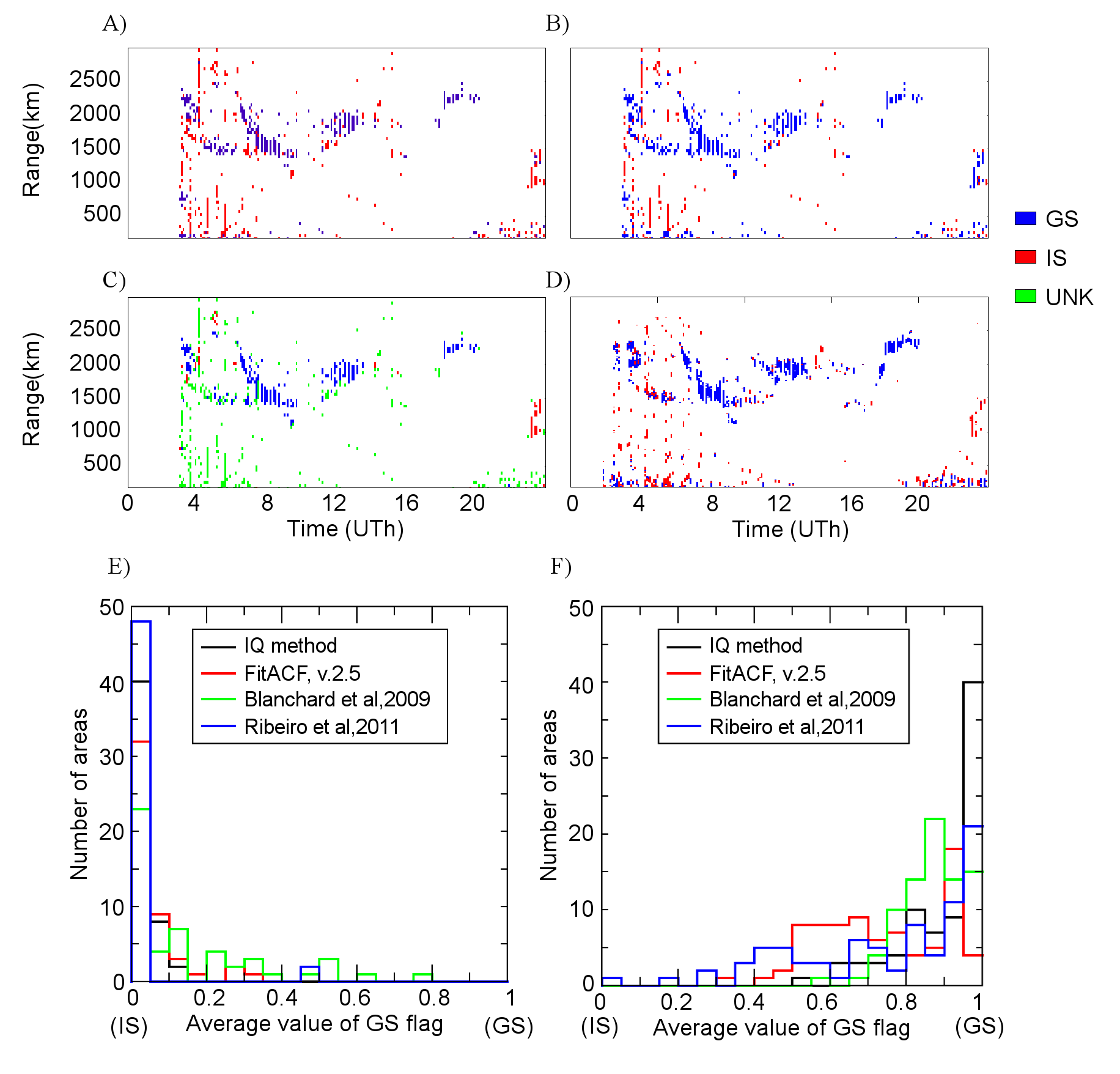}
\caption[test]{ A) - Identification by \cite{FitACF_25};
B) - Identification by \cite{Blanchard_2009};
C) - Identification by \cite{Ribeiro_2011};
D) - Identification by our IQ method;
Blue corresponds to GS, red to IS, green to unknown signals;
E-F) The distribution of mean signal type over regions:
E) - results of processing IS data; F) - results of processing GS data;
Black, red, green and blue lines correspond to IQ method,\cite{FitACF_25},\cite{Blanchard_2009} and \cite{Ribeiro_2011} methods correspondingly.
}
\label{fig:FIG12}
\end{figure}

In the analysis, each case of scattering was analyzed by each of the four algorithms.
Example of signal type identification is shown in Fig.\ref{fig:FIG12}A-D.
The distribution of the GS flags for different signal types is shown in Fig.\ref{fig:FIG12}E-F 
(Fig.\ref{fig:FIG12}E - IS signals only, Fig.\ref{fig:FIG12}E - GS signals only). 
The distribution of the errors is not the same for different methods. From Fig.\ref{fig:FIG12}E-F one can 
see that \cite{Ribeiro_2011} looks the most effective for detecting IS and GS signals
(over the data it can identify).

The proposed IQ-algorithm showed the smallest value of the total error (the average total error is $13.3\%$).
The error of GS incorrectly defined as IS is $11.2\%$, and the erroneous definition of IS as GS is $2.1\%$. 
The \cite{FitACF_25} algorithm showed a generally worse accuracy in both cases ($26.3\%$ and $6.6\%$,  respectively).
Using the \cite{Blanchard_2009} algorithm produce errors of $13.1\%$ and $15.6\%$, respectively.
The smallest total error ($25.6\%$) of the three traditional algorithms was produced by the \cite{Ribeiro_2011} algorithm, 
but it did not process a large number of scattered signals ($23.4\%$ GS and $54.9\%$ IS, see Fig.\ref{fig:FIG12}C) at all, which was not considered an error in our estimates.

\section{Conclusion}

We made the analysis of the fine structure of decameter signals scattered
by irregularities of the Earth's surface and field-aligned ionospheric
irregularities. The software of the EKB ISTP SB RAS radar was substantially
modernized to carry out such an analysis with a high sampling
frequency. Large number of experimental data with an increased
sampling frequency has been obtained. 

As a result of the experimental
data analysis it is shown that signals scattered by both mechanisms
have a specific phase structure and a nonzero lifetime. This allows
us to interpret them as a superposition of elementary responses with a finite lifetime 
and specific amplitude-phase shape. An algorithm is constructed for estimating the 
shape of the elementary response. 
A unified empirical model of elementary response is developed, that allows us 
to describe both types of elementary response and to determine their characteristics 
- the lifetime, the duration of the right and left edges. 
Statistical features of elementary responses in GS and IS signals are estimated and presented. 
Differences in the properties of elementary responses in GS and IS signals are found: 
the different duration of the right front of the signal and different
lifetimes. 

Within the Bayesian inference approach and based on the analysis of the full waveform,
a technique for the optimal identification of GS and IS signals was constructed. The technique (IQ method)
works without using traditional SuperDARN algorithm for estimating scattered signal parameters (FitACF) 
- spectral width and Doppler drift velocity. It is based on the analysis of IQ components of the signal only.

The weakness of the method is the need for a relatively high SNR ($>2$), at smaller SNR it is unstable and should not be used. 
Another weakness of the method is its non-local character. 
It can incorrectly identify the type of signals separated by a distance of less than 200~km, 
identifying the type of signal by a more powerful signal. In addition, within the spatial 
resolution (45~km), it may incorrectly identify the GS signal as IS signal at the boundary 
between lower ranges GS signal and higher ranges noise.

The effectiveness of the method is estimated based on EKB radar data at SNR$>2$.
It is shown that using this technique at EKB radar at SNR$>2$ produces
less total error, than traditional algorithms \cite{FitACF_25,Blanchard_2009,Ribeiro_2011}.
The total error (defined as the total ratio of incorrectly defined points in all areas to 
their total number) produced by our IQ-algorithm over the selected data was $13.3\%$.
\cite{FitACF_25} algorithm produces total error of $32.9\%$, \cite{Blanchard_2009} method produces total error about $28.7\%$, 
\cite{Ribeiro_2011} algorithm produces total error about $25.6\%$ (over the data, marked definitely as GS or IS). 

So over test dataset of EKB radar data with SNR$>2$ our IQ algorithm provides noticeably higher accuracy in determining the 
type of signal than other methods.
To identify signal types at low SNR ($<2$) or at 45~km boundary with noise, one can use the 
traditional local methods \cite{FitACF_25,Blanchard_2009} instead of this technique.

The work was supported by the RFBR grant \# 16-05-01006a. In the paper EKB ISTP SB RAS radar data were used, 
the radar functioning is supported by the FSR program II.12. The data
of EKB radar is a property of ISTP SB RAS, contact Oleg Berngardt (berng@iszf.irk.ru).


%
%





\end{document}